\documentclass[12pt]{iopart}

\usepackage{amssymb}
\usepackage{graphicx}

\begin{document}

\title[A novel Hirota bilinear approach to $N=2$ supersymmetric equations]{A novel Hirota bilinear approach to $N=2$ supersymmetric equations}

\author{Laurent DELISLE}

\address{ECE Engineering school, 10 rue Sextius Michel, 75015 Paris, France}
\ead{laurent.delisle@ece.fr}


\begin{abstract}
This article presents a novel application of the Hirota bilinear formalism to the $N=2$ supersymmetric KdV and Burgers equations. This new approach avoids splitting $N=2$ equations into two $N=1$ equations. We use the super Bell polynomials to obtain bilinear representations and present multi-soliton solutions.
\end{abstract}

%
%
%
%
%

\section{Introduction}

The study of exact solutions of completely integrable supersymmetric systems is of current interest in modern mathematical physics research. In particular, the $N=$2 supersymmetric extension of the Korteweg-de Vries (KdV) equation \cite{Labelle} has been largely studied in terms of integrability conditions, exact solutions and symmetry group structures \cite{Ayari,Ghosh,Hussin,Zhang,Delisle,Delisle1,Hussin1,Liu,Tian,Delisle2}. The equation is described by a bosonic superfield $A$ defined on the superspace $\mathbb{R}^{2\vert 2}$ \cite{Cornwell} of local coordinates $(x,t,\theta_1,\theta_2)$. The variables $(x,t)$ are the usual space-time coordinates and $(\theta_1,\theta_2)$ are real Grassmann coordinates satisfying the usual anti-commutation rules
\begin{equation}
\theta_1\theta_2+\theta_2\theta_1=0\quad \mbox{and}\quad  \theta_1^2=\theta_2^2=0.
\end{equation}

The superfield $A$ satisfies the Labelle-Mathieu $N=2$ supersymmetric extension of the KdV equation \cite{Labelle}
\begin{equation}
A_t=(-A_{xx}+(a+2)AD_1D_2A+(a-1)(D_1A)(D_2A)+aA^3)_x,
\label{SKdV}
\end{equation}
where $a\in\mathbb{R}$ is a real parameter, $D_1$, $D_2$ are the covariant derivatives defined as
\begin{equation}
D_{\mu}=\partial_{\theta_{\mu}}+\theta_{\mu} \partial_x, \quad \mu=1,2
\end{equation}
and satisfy $D_1^2=D_2^2=\partial_x$. The bosonic superfield $A$ can be decomposed \cite{Cornwell} using a Taylor expansion around $(x,t,0,0)$ as
\begin{equation}
A(x,t,\theta_1,\theta_2)=u(x,t)+\theta_1\xi_1(x,t)+\theta_2\xi_2(x,t)+\theta_1\theta_2 v(x,t),
\label{fieldA}
\end{equation}
where $u$ and $v$ are complex-valued even functions and $\xi_1$, $\xi_2$ are complex-valued odd functions.

Labelle and Mathieu \cite{Labelle} showed that equation (\ref{SKdV}) is completely integrable for the special values $a=-2,1,4$. This fact suggests that, for these special values, the supersymmetric KdV equation possesses multi-soliton solutions \cite{Ayari,Ghosh,Hussin,Zhang,Delisle,Delisle1,Delisle2}. An algebraic direct method of obtaining such solutions is described by the Hirota bilinear formalism \cite{Hirota}. This method was used numerous time for non-supersymmetric and $N=1$ supersymmetric integrable equations to construct soliton and similarity solutions, Bäcklund and Darboux transformations and to obtain integrability conditions \cite{Ablowitz,Hirota,Carstea,Delisle3}. Carstea has adapted this formalism to $N=1$ supersymmetric extensions \cite{Carstea} such as the KdV, modified KdV and Sine-Gordon equations. The generalization of this formalism to $N=2$ extensions has been confronted to numerous difficulties. Zhang and \textit{al.} \cite{Zhang} used the strategy of decomposing equation (\ref{SKdV}) into two $N=1$ equations for which the Hirota formalism is well adapted. The way of achieving this is to re-write the bosonic superfield $A$ given in (\ref{fieldA}) as
\begin{equation}
A(x,t,\theta_1,\theta_2)=A_0(x,t,\theta_1)+\theta_2 \Xi(x,t,\theta_1),
\label{SplitA}
\end{equation}
where $A_0(x,t,\theta_1)=u(x,t)+\theta_1\xi_1(x,t)$ and $\Xi(x,t,\theta_1)=\xi_2(x,t)-\theta_1v(x,t)$ are, respectively, even and odd superfields of $(x,t,\theta_1)\in\mathbb{R}^{2\vert 1}$. In order to use the bilinear Hirota formalism \cite{Ablowitz,Hirota}, we have to re-write the superfields $A_0$ and $\Xi$ in terms of dimensionless bosonic superfields. This is done by dimensional analysis following the fact that equation (\ref{SKdV}) is invariant under the dilatation vector superfield \cite{Ayari}
\begin{equation}
x\partial_x+3t\partial_t+\frac12 \theta_1\partial_{\theta_1}+\frac12\theta_2\partial_{\theta_2}-A\partial_A.
\end{equation}
This vector superfield shows that, under the transformation
\begin{equation}
(x,t,\theta_1,\theta_2,A)\rightarrow \left(\lambda x,\lambda^3t,\lambda^{\frac12}\theta_1,\lambda^{\frac12}\theta_2,\lambda^{-1}A\right),
\end{equation}
equation (\ref{SKdV}) is invariant, where $\lambda$ is a free even parameter. Under these transformations, we deduce the dimension of these quantities
\begin{equation}
[A]=-1, \quad [\partial_x]=-1,\quad [D_1]=-\frac12
\end{equation}
and this allows us to re-write the superfields $A_0$ and $\Xi$ as
\begin{equation}
A_0=B_x,\quad \Xi=D_1P_x,
\label{dimensionless}
\end{equation}
where $B=B(x,t,\theta_1)$ and $P=P(x,t,\theta_1)$ are dimensionless bosonic superfields. Introducing the superfield $A$ as written in (\ref{SplitA}) together with the relations (\ref{dimensionless}) into equation (\ref{SKdV}) gives two $N=1$ supersymmetric equations
\begin{eqnarray}
B_t&=&-B_{xxx}+(a+2)B_xP_{xx}+(a-1)(D_1B_x)(D_1P_x)+aB_x^3,\\
D_1P_t&=&-D_1P_{xxx}+3P_{xx}D_1P_x-(a+2)B_xD_1B_{xx}\\&\phantom{=}&+(1-a)B_{xx}D_1B_x+3aB_x^2D_1P_x\nonumber.
\end{eqnarray}
The novel approach in this article is to avoid splitting the $N=2$ supersymmetric equation into these two $N=1$ supersymmetric equations. The idea is to directly apply the Hirota bilinear formalism to equation (\ref{SKdV}). The relations (\ref{dimensionless}) will nevertheless inspire us. We thus observe that our superfield $A$ written as in (\ref{SplitA}) together with the relations (\ref{dimensionless}) reads as $A=C_x$ where $C(x,t,\theta_1,\theta_2)=B(x,t,\theta_1)+\theta_2D_1P(x,t,\theta_1)$. After one integration with respect to $x$, equation (\ref{SKdV}) as the following form

\begin{equation}
C_t=-C_{xxx}+(a+2)C_xD_1D_2C_x+(a-1)(D_1C_x)(D_2C_x)+aC_x^3.
\label{potentialKdV}
\end{equation}
In this article, we will consider the special case $a=1$ to illustrate our method. For $a\neq 1$, a new problem arises from the product of the odd superfields $D_1 C_x$ and $D_2 C_x$. This is an ongoing project and have been partially answered in the $a=4$ case \cite{Zhang,Delisle2}.

\section{The supersymmetric KdV equation with $a=1$} 

Let us consider equation (\ref{potentialKdV}) with $a=1$:
\begin{equation}
C_t=-C_{xxx}+3C_xD_1D_2C_x+C_x^3.
\label{KdVa=1}
\end{equation}
The difficulties involved in using the Hirota bilinear formalism in the $N=2$ supersymmetric context reside in the terms that are written using the covariant derivatives. In the case $a=1$, the difficult term is $D_1D_2C_x$. However, the superfield $C$ was written as $C=B+\theta_2D_1P$ which implies that
\begin{equation}
D_1D_2C_x=P_{xx}-\theta_2D_1B_{xx}=W_{xx},
\label{constraint}
\end{equation}
where $W(x,t,\theta_1,\theta_2)=P(x,t,\theta_1)-\theta_2D_1B(x,t,\theta_1)$. Equation (\ref{KdVa=1}) can thus be re-written in a "non-supersymmetric way" as
\begin{equation}
C_t=-C_{xxx}+3C_xW_{xx}+C_x^3,
\label{bosonicKdVa=1}
\end{equation}
together with the constraint (\ref{constraint}) which reads as $W_{xx}=D_1D_2C_x$. To obtain a bilinear Hirota representation of equations (\ref{bosonicKdVa=1}) and (\ref{constraint}), we will use the supersymmetric extension of the Bell polynomials. In what follows, we introduce the one-variable and binary Bell polynomials and its relation with the Hirota formalism in the supersymmetric context \cite{Fan}.

The one-variable Bell polynomials $Y$ are defined as
\begin{equation}
Y_{k_xx,k_tt,k_1\theta_1,k_2\theta_2}(f)=e^{-f}\partial_x^{k_x}\partial_t^{k_t}D_1^{k_1}D_2^{k_2}e^f,
\end{equation}
where $k_{\mu}$ are positive integers and $f=f(x,t,\theta_1,\theta_2)$ is an even superfield. Using the polynomials $Y$, we define the binary Bell polynomials $\mathcal{Y}$ as
\begin{equation}
\mathcal{Y}_{k_xx,k_tt,k_1\theta_1,k_2\theta_2}(\omega_1,\omega_2)=Y_{k_xx,k_tt,k_1\theta_1,k_2\theta_2}(f_{\tilde{k}_xx\tilde{k}_tt\tilde{k}_1\theta_1\tilde{k}_2\theta_2}),
\end{equation}
where the different derivatives of $f$ are replaced by the superfields $\omega_1$ and $\omega_2$ following the procedure
\begin{equation}
f_{\tilde{k}_xx\tilde{k}_tt\tilde{k}_1\theta_1\tilde{k}_2\theta_2} = \left\{
    \begin{array}{ll}
        \omega_{1,\tilde{k}_xx\tilde{k}_tt\tilde{k}_1\theta_1\tilde{k}_2\theta_2} & \mbox{if } \tilde{k}_x+\tilde{k}_t+\tilde{k}_1+\tilde{k}_2 \mbox{is odd}, \\
       \omega_{2,\tilde{k}_xx\tilde{k}_tt\tilde{k}_1\theta_1\tilde{k}_2\theta_2}  & \mbox{otherwise.}
    \end{array}
\right.
\end{equation}
Note that the following notation is used: $f_{\tilde{k}_xx\tilde{k}_tt\tilde{k}_1\theta_1\tilde{k}_2\theta_2}=\partial_x^{\tilde{k}_x}\partial_t^{\tilde{k}_t}D_1^{\tilde{k}_1}D_2^{\tilde{k}_2}f$. The link with the Hirota bilinear formalism is given by
\begin{eqnarray}
\mathcal{Y}_{k_xx,k_tt,k_1\theta_1,k_2\theta_2}\left(\omega_1=\ln\left(\frac{f}{g}\right),\omega_2=\ln (fg)\right)=(fg)^{-1}\mathcal{S}_1^{k_1}\mathcal{S}_2^{k_2}\mathcal{D}_x^{k_x}\mathcal{D}_t^{k_t}(f. g),\nonumber\\
\phantom{=}
\label{HirotaBell}
\end{eqnarray}
where the Hirota derivative is defined as
\begin{eqnarray}
\mathcal{S}_{\mu}^k\mathcal{D}_{x}^j(f.g)=(D_{\mu}-D'_{\mu})^k(\partial_{x}-\partial_{x'})^jf(x,\theta)g(x',\theta')\vert_{x'=x,\theta_{\mu}'=\theta_{\mu}}, \quad \mu=1,2.\nonumber\\
\phantom{=}
\end{eqnarray}
In particular, taking $g=f$ in equation (\ref{HirotaBell}), we get
\begin{eqnarray}
\mathcal{P}_{k_xx,k_tt,k_1\theta_1,k_2\theta_2}(f)&=&\mathcal{Y}_{k_xx,k_tt,k_1\theta_1,k_2\theta_2}\left(\omega_1=0,\omega_2=2\ln (f)\right))\nonumber\\
&=&f^{-2}\mathcal{S}_1^{k_1}\mathcal{S}_2^{k_2}\mathcal{D}_x^{k_x}\mathcal{D}_t^{k_t}(f. f).\label{specialBell}
\end{eqnarray}
Making use of the Bell polynomials, equation (\ref{bosonicKdVa=1}) can be written as
\begin{equation}
\mathcal{Y}_t(iC,-W)+\mathcal{Y}_{3x}(iC,-W)=0.
\end{equation}
Taking
\begin{equation}
iC=\ln\left(\frac{F}{G}\right)\quad \mbox{and} \quad -W=\ln\left(FG\right)
\label{changeofvar}
\end{equation}
with $F=F(x,t,\theta_1,\theta_2)$, $G=G(x,t,\theta_1,\theta_2)$ bosonic superfields, we get, using relation (\ref{HirotaBell}), the Hirota representation of equation (\ref{bosonicKdVa=1}):
\begin{equation}
(\mathcal{D}_t+\mathcal{D}_x^3)(F.G)=0.
\end{equation}
Now, the problem (and "novelty") resides in finding a Hirota representation of constraint (\ref{constraint}). This subject is one of the novelty of this article and is a crucial part in avoiding splitting a $N=2$ supersymmetric equation into two $N=1$ equations. After one integration with respect to $x$ and using $D_1^2=\partial_x$, relation (\ref{constraint}) which arises from dimension analysis takes the following form
\begin{equation}
D_1^2W=D_1D_2C.
\end{equation}
However, the change of variables (\ref{changeofvar}) allows us to re-write the above constraint as
\begin{equation}
D_1(D_1-iD_2)(2\ln F)+D_1(D_1+iD_2)(2\ln G)=0.
\end{equation}
This equation as a Hirota bilinear representation using the Bell polynomials (\ref{specialBell}) as
\begin{equation}
F^{-2}\mathcal{S}_1(\mathcal{S}_1-i\mathcal{S}_2)(F.F)+G^{-2}\mathcal{S}_1(\mathcal{S}_1+i\mathcal{S}_2)(G.G)=0.
\end{equation}
Hence, we get a $N=2$ Hirota bilinear representation of the supersymmetric equation (\ref{SKdV}) for $a=1$ given as
\begin{equation}
\left\{\begin{array}{ll}
(\mathcal{D}_t+\mathcal{D}_x^3)(F.G)&=0,\\
\mathcal{S}_1(\mathcal{S}_1-i\mathcal{S}_2)(F.F)&=0,\\
\mathcal{S}_1(\mathcal{S}_1+i\mathcal{S}_2)(G.G)&=0.
\end{array}
\right.
\label{HirotaFinal}
\end{equation}
This result is new and allows $N=2$ Hirota representation of supersymmetric equation in terms of superfields in $\mathbb{R}^{2\vert 2}$ superspace. As of now, Hirota representation of $N=2$ supersymmetric equations where in terms of superfields in $\mathbb{R}^{2\vert 1}$ superspace \cite{Zhang,Delisle2}.

\section{Soliton solutions of the supersymmetric KdV equation with $a=1$}
To find soliton solutions of the supersymmetric KdV equation with $a=1$, we have to solve the system of Hirota equations (\ref{HirotaFinal}). We will do this by supposing that the superfields $F$ and $G$ are written as sums of exponentials \cite{Ablowitz}. Adding exponential is equivalent to adding a solitary wave. Meaning that if $F$ and $G$ are composed of $N$ independent exponential functions, then the solution of equation (\ref{SKdV}) will correspond to a $N$-soliton solution.
\subsection{The solitary wave solution}

Let us suppose that a solution of system (\ref{HirotaFinal}) is given by
\begin{equation}
F=a_{0,F}+a_{1,F} e^{\Lambda_F} \quad \mbox{and}\quad G=a_{0,G}+a_{1,G} e^{\Lambda_G},
\label{onesolitonFG}
\end{equation}
where $a_{\mu,F}$, $a_{\mu,G}$ are even real constants and $\Lambda_{\mu}$ are even superfields which can be written as
\begin{equation}
\Lambda_{\mu}=\kappa_{\mu}x+\omega_{\mu}t+\theta_1\eta_{1,\mu}+\theta_2\eta_{2,\mu}+\theta_1\theta_2m_{12,\mu},
\end{equation}
for $\mu=F,G$, where $\kappa_{\mu}$, $\omega_{\mu}$, $m_{12,\mu}$ are even real constants and $\eta_{1,\mu}$, $\eta_{2,\mu}$ are odd real constants to be determined. In what follows, we consider a special case for the solitary wave solution :

\medskip

\noindent In order to simplify the expressions of the components of the solution $A$, we take $a_{0,F}=a_{0,G}=1$ and $a_{1,F}a_{1,G}\neq 0$: 
Introducing (\ref{onesolitonFG}) into the system (\ref{HirotaFinal}) imposes the following forms :
\begin{eqnarray}
\Lambda_F&=&\kappa x-\kappa^3t+(\theta_1-i\theta_2)\eta_F+i\theta_1\theta_2\kappa,\\
\Lambda_G&=&\kappa x-\kappa^3t+(\theta_1+i\theta_2)\eta_G-i\theta_1\theta_2\kappa.
\end{eqnarray}
In this case, the even components of the solution superfield $A$ as shown in (\ref{fieldA}) are given as
\begin{eqnarray}
u(x,t)&=&i\frac{(a_{1,G}-a_{1,F})\kappa e^{\lambda}}{(1+a_{1,G} e^{\lambda})(1+a_{1,F} e^{\lambda})},\\ 
v(x,t)&=&\frac{\kappa^2e^{\lambda}\left(4a_{1,F}a_{1,G}e^{\lambda}+(a_{1,F}+a_{1,G})(1+a_{1,F}a_{1,G}e^{2\lambda})\right)}{(1+a_{1,F}e^{\lambda})^2(1+a_{1,G}e^{\lambda})^2},
\end{eqnarray}
where $\lambda=\kappa x-\kappa^3 t$. In order to simplify those expressions, we take, without lost of generality, $a_{1,G}=1$ and $a_{1,F}=-1$. We get
\begin{equation}
u(x,t)=-\frac{i\kappa}{\sinh(\lambda)}\quad \mbox{and}\quad v(x,t)=u^2(x,t).
\end{equation}
For the odd components of the solution superfield $A$, we get
\begin{eqnarray}
\xi_1&=&\frac{i\kappa e^{\lambda}\left((\eta_G+\eta_F)(1+e^{2\lambda})+2(\eta_F-\eta_G)e^{\lambda}\right)}{(1-e^{2\lambda})^2},\\
\xi_2&=&\frac{\kappa e^{\lambda}\left((\eta_F-\eta_G)(1+e^{2\lambda})+2(\eta_F+\eta_G)e^{\lambda}\right)}{(1-e^{2\lambda})^2}.
\end{eqnarray}
The bosonic components of the solutions have been obtained in \cite{Zhang}. The main difference lies in the fermionic components, which allows two odd constants $\eta_F$ and $\eta_G$. This particularity allows one to generalize the solutions obtained in \cite{Zhang}.\\

To simplify those expressions, we take $\eta_F=\eta_G=\eta$. We get, in this case,
\begin{equation}
\xi_1(x,t)=\frac{i\kappa\eta}{\sinh(\lambda)\tanh(\lambda)}, \quad \xi_2(x,t)=\frac{\kappa\eta}{\sinh^2(\lambda)}.
\end{equation}

An other interesting choice is $a_{1,F}=a_{1,G}=1$ together with $\eta_F=\eta_G=\eta$. In this case, we get for the even components of $A$
\begin{equation}
u(x,t)=0,\quad v(x,t)=\frac{\kappa^2}{2 \cosh^2\left(\frac{\lambda}{2}\right)},
\end{equation}
and for the odd components:
\begin{equation}
\xi_1(x,t)=0,\quad \xi_2(x,t)=-\frac{\kappa\eta}{2\cosh^2(\frac{\lambda}{2})}.
\end{equation}

\subsection{The $2$-soliton solution}
This subsection exhibits different types of 2-soliton solution. Let us suppose that a solution of system (\ref{HirotaFinal}) is given by
\begin{eqnarray}
F&=&a_{0,F}+a_{1,F}e^{\Lambda_{1,F}}+a_{2,F}e^{\Lambda_{2,F}}+a_{12,F}e^{\Lambda_{1,F}+\Lambda_{2,F}}\label{fieldF},\\
G&=&a_{0,G}+a_{1,G}e^{\Lambda_{1,G}}+a_{2,G}e^{\Lambda_{2,G}}+a_{12,G}e^{\Lambda_{1,G}+\Lambda_{2,G}}\label{fieldG},
\end{eqnarray}
where $\Lambda_{\mu,\nu}$ are even superfields given as
\begin{eqnarray}
\Lambda_{\mu,F}&=&\lambda_{\mu}+(\theta_1-i\theta_2)\eta_{\mu,F}+i\theta_1\theta_2\kappa_{\mu},\\
\Lambda_{\mu,G}&=&\lambda_{\mu}+(\theta_1+i\theta_2)\eta_{\mu,G}-i\theta_1\theta_2\kappa_{\mu},
\end{eqnarray}
and $\lambda_{\mu}=\kappa_{\mu}x-\kappa_{\mu}^3t$ for $\mu=1,2$ and the $a$'s and $b$'s are bosonic constants. These even superfields satisfy the following relations, which are helpful in calculations,
\begin{equation}
(D_1-iD_2)\Lambda_{\mu,F}=0 \quad \mbox{and} \quad (D_1+iD_2)\Lambda_{\mu,G}=0.
\end{equation}
Indeed, we deduce $D_1D_2\Lambda_{\mu,F}=-i\kappa_{\mu}=-D_1D_2\Lambda_{\mu,G}$ and $D_1\Lambda_{\mu,F}D_2\Lambda_{\mu,F}=D_1\Lambda_{\mu,G}D_2\Lambda_{\mu,G}=0$.

Introducing expressions (\ref{fieldF}) and (\ref{fieldG}) into system (\ref{HirotaFinal}), we get the following set of constraints on the bosonic constants:
\begin{equation}
a_{0,F}a_{12,G}=a_{1,F}a_{2,G}=a_{2,F}a_{1,G}=a_{12,F}a_{0,G}=0.
\label{constraints1}
\end{equation}
Below, we present two special cases for the $2$-soliton solution.

\begin{enumerate}
\item \textbf{Case 1}: We take, for example, $a_{0,F}=a_{12,F}=a_{1,G}=a_{2,G}=0$. To simplify the expressions of the components of the superfield $A$, we make the choice $a_{1,F}=a_{2,F}=a_{0,G}=a_{12,G}=1$. In this case, the superfields $F$ and $G$ have the following forms:
\begin{equation}
F=e^{\Lambda_{1,F}}+e^{\Lambda_{2,F}} \quad \mbox{and} \quad G=1+e^{\Lambda_{1,G}+\Lambda_{2,G}}.
\end{equation}

 The even components $u$ and $v$ of the superfield $A$ as the following forms:
\begin{equation}
u(x,t)=\frac{i}{2}\times \left( \frac{\kappa_2\sinh(\lambda_1)+\kappa_1\sinh(\lambda_2)}{\cosh\left(\frac{\lambda_1-\lambda_2}{2}\right)\cosh\left(\frac{\lambda_1+\lambda_2}{2}\right)}\right)
\end{equation}
and
\begin{equation}
v(x,t)=\frac14 \left(\frac{(\kappa_1-\kappa_2)^2}{\cosh^2\left(\frac{\lambda_1-\lambda_2}{2}\right)}+\frac{(\kappa_1+\kappa_2)^2}{\cosh^2\left(\frac{\lambda_1+\lambda_2}{2}\right)}\right).
\end{equation}
In \textbf{Figure 1.}, we exhibit these functions for $\kappa_1=1$, $\kappa_2=2$ and for times $t=-2,-1,0,2$. The graph of $\Im (u)$ is represented by a full black curve and the graph of function $v$ by a dotted curve.

\begin{figure}
\begin{center}
\includegraphics[scale=0.5,height=3cm,width=6cm]{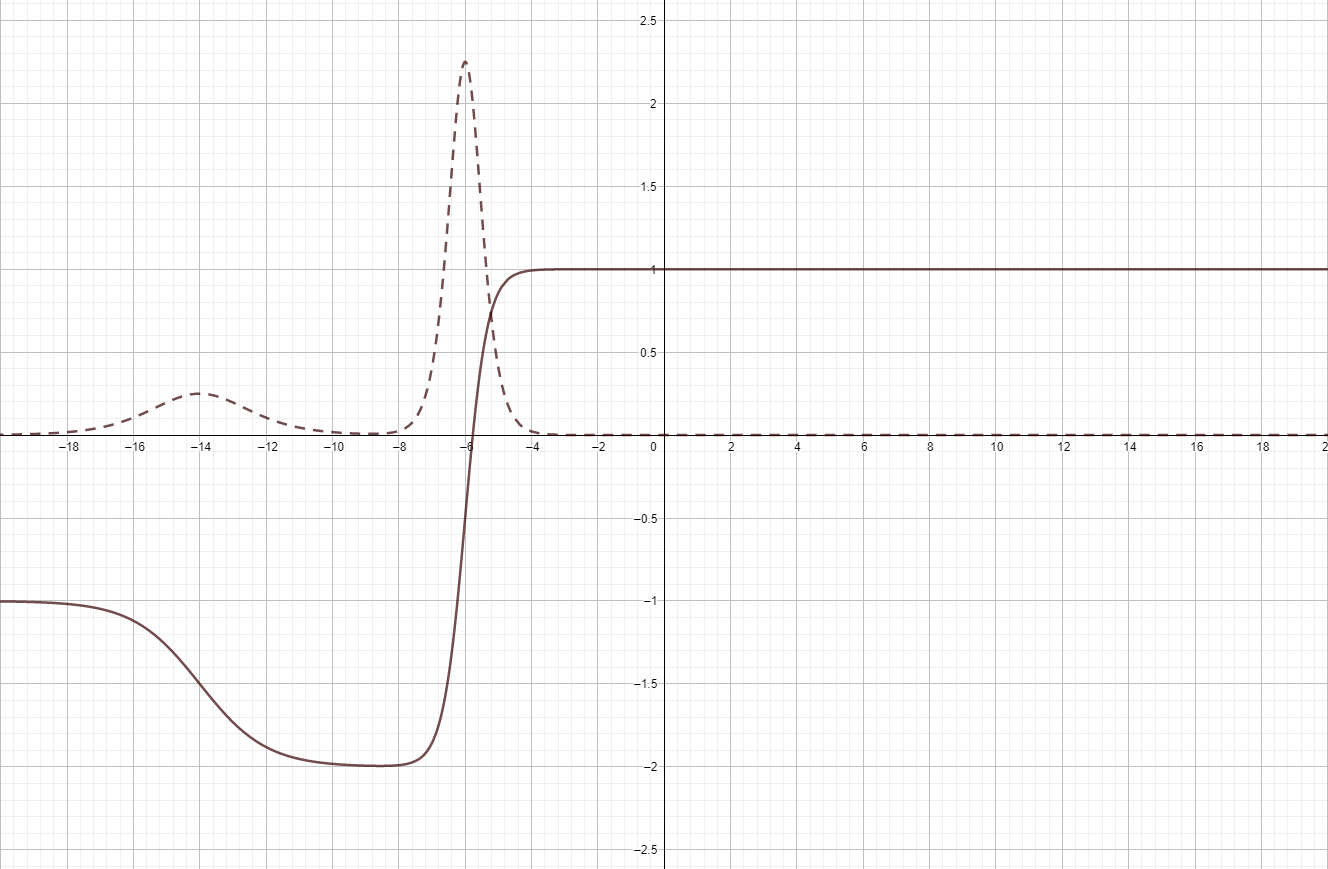}
\includegraphics[scale=0.5,height=3cm,width=6cm]{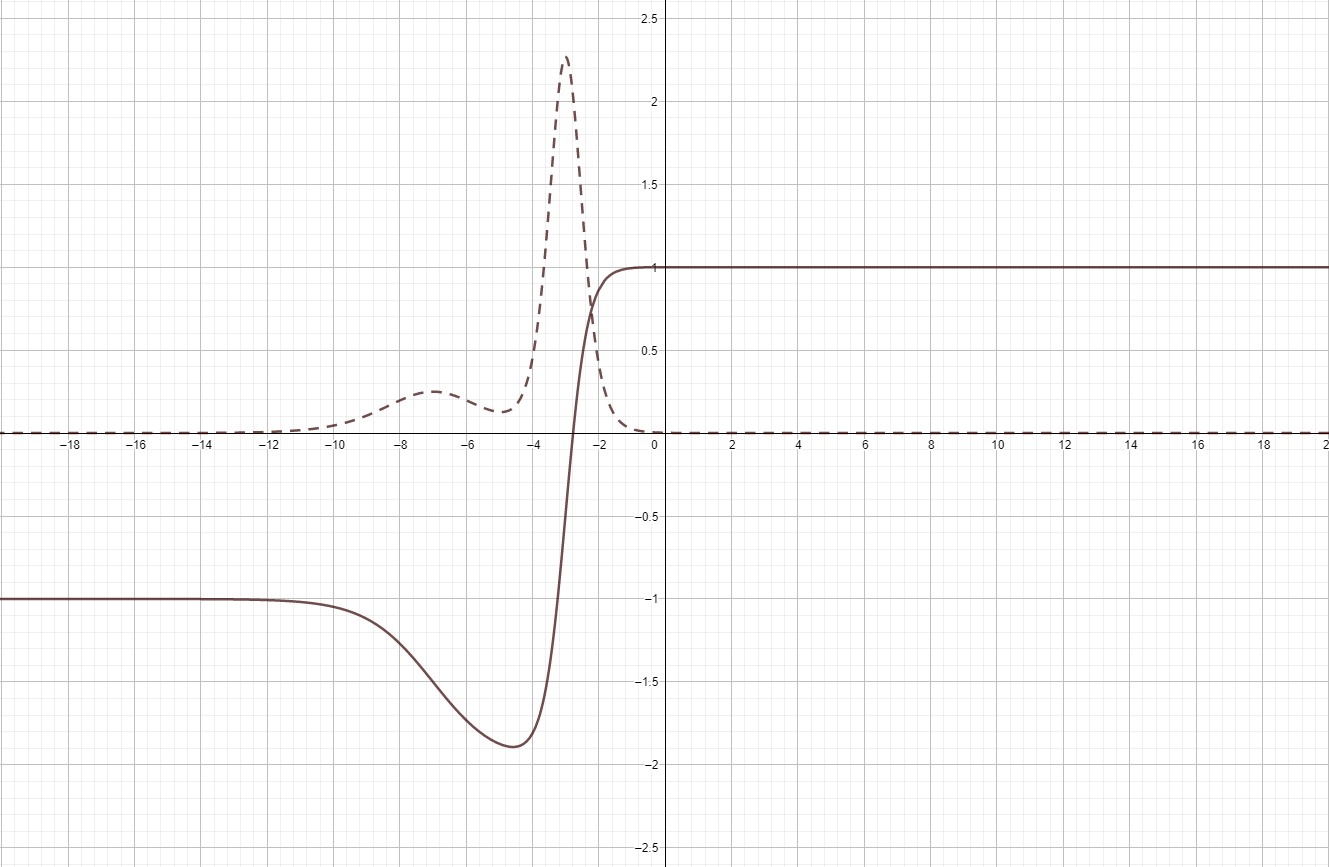}\\
\includegraphics[scale=0.5,height=3cm,width=6cm]{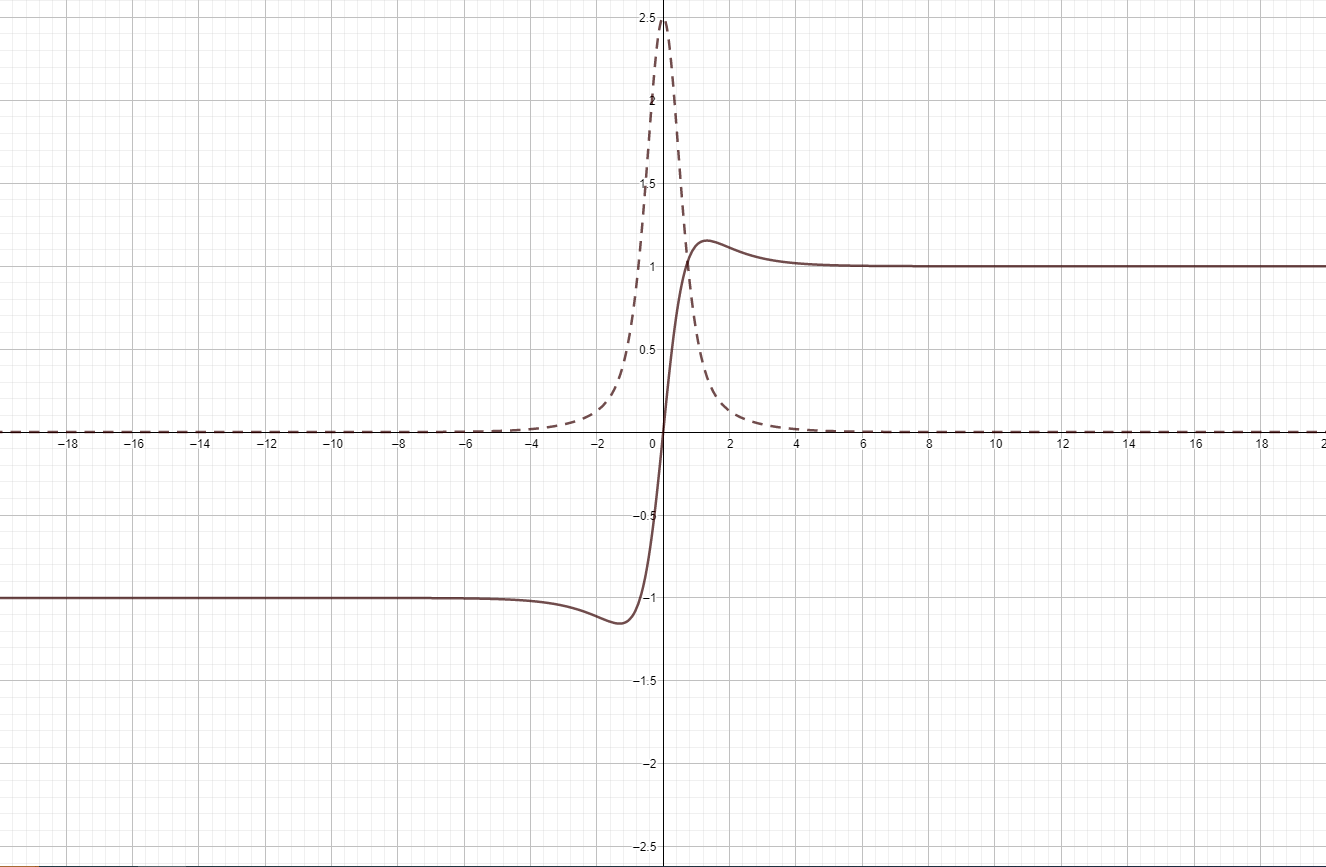}
\includegraphics[scale=0.5,height=3cm,width=6cm]{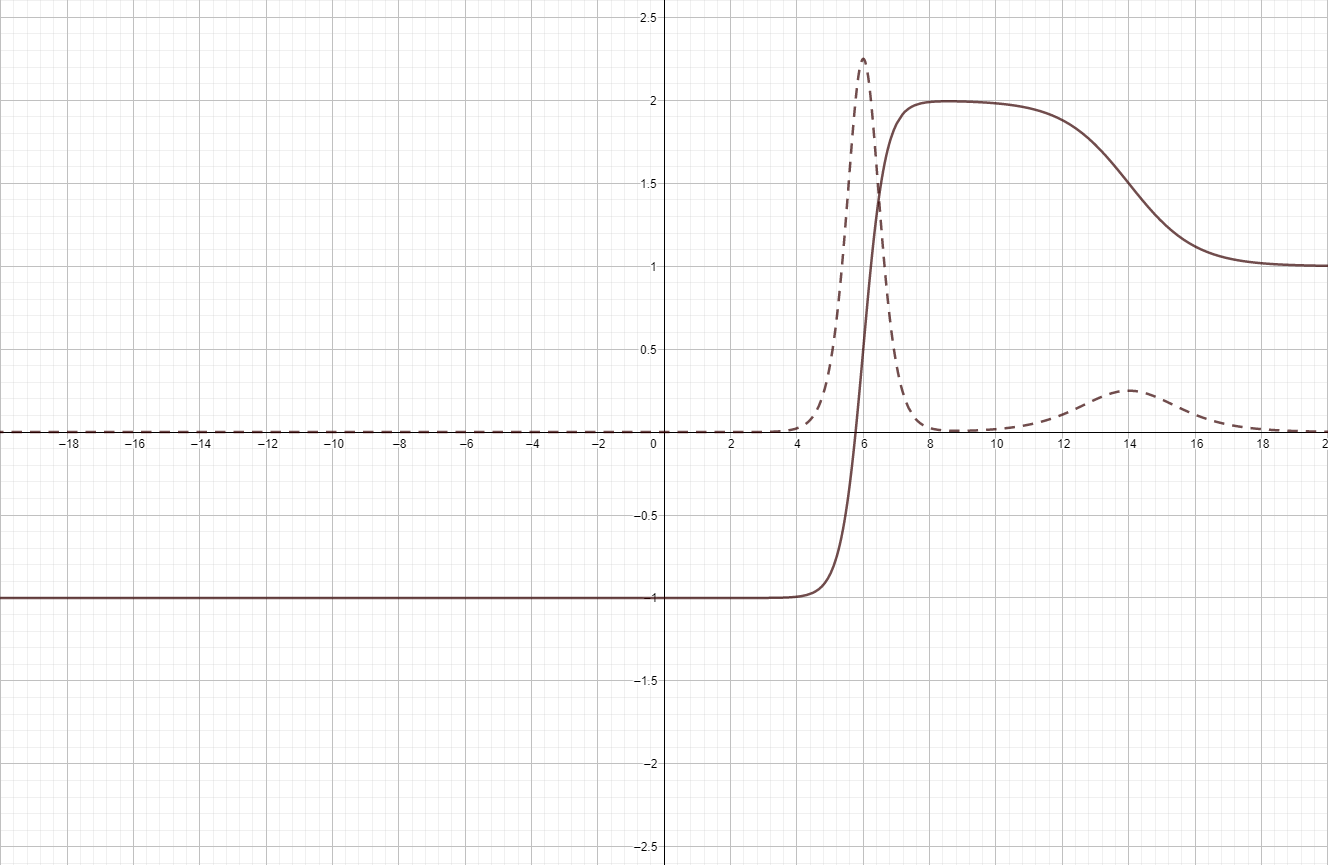}
\caption{ The functions $\Im(u)$ and $v$ for $\kappa_1=1$, $\kappa_2=2$ and, from left to right, $t=-2,-1,0,2$.}
\end{center}
\end{figure}

The behavior of these soliton solutions is quite different from the soliton behavior obtained in \cite{Zhang}. Indeed, in the latest, the authors retrieve a 2-soliton solution which blows up for certain values of $x$ and $t$. This shows, that, we have obtained new type of multi-soliton solutions of the supersymmetric KdV equation.
\item \textbf{Case 2:} We take $a_{0,F}=a_{12,F}=a_{1,G}=a_{2,G}=a_{0,G}=0$ and $a_{1,F}=a_{2,F}=a_{12,G}=1$. In this case, the superfields $F$ and $G$ have the following forms:
\begin{equation}
F=e^{\Lambda_{1,F}}+e^{\Lambda_{2,F}} \quad \mbox{and} \quad G=e^{\Lambda_{1,G}+\Lambda_{2,G}}.
\end{equation}
In this case, we impose $a_{0,G}=0$ which is unnecessary in order for the constraints (\ref{constraints1}) to be satisfied, but will be helpful in the generalisation of a $N$-soliton solution for $N\geq 2$.\\
The even components $u$ and $v$ of the superfield $A$ as the following forms: 
\begin{equation}
u(x,t)=\frac{i}{2}\times \left(\kappa_1+\kappa_2+(\kappa_2-\kappa_1)\tanh\left(\frac{\lambda_1-\lambda_2}{2}\right)\right)
\end{equation}
and
\begin{equation}
v(x,t)=\frac{1}{2}\times \left(\kappa_1^2+\kappa_2^2+(\kappa_1^2-\kappa_2^2)\tanh\left(\frac{\lambda_1-\lambda_2}{2}\right)\right).
\end{equation}
\end{enumerate}

\subsection{The $3$-soliton solution and generalization}

For the $3$-soliton solution, we consider

\begin{equation}
F=e^{\Lambda_{1,F}+\Lambda_{2,F}}+e^{\Lambda_{1,F}+\Lambda_{3,F}}+e^{\Lambda_{2,F}+\Lambda_{3,F}} \quad \mbox{and}\quad G=e^{\Lambda_{1,G}+\Lambda_{2,G}+\Lambda_{3,G}},
\end{equation}
where $\Lambda_{\mu,F}$ and $\Lambda_{\mu,G}$ are given in previous subsections and $a_0$ and $a_3$ are arbitrary bosonic constants. The difficulty (and novelty) here, which was absent in the classical and $N=1$ supersymmetric cases, is that $\Lambda_{\mu,F}\neq\Lambda_{\mu,G}$. This limits the form of $F$ and $G$ so that they satisfy the bilinear equation $(\mathcal{D}_t+\mathcal{D}_x^3)(F . G)=0$.

We can generalize the $3$-soliton solution to a $N$-soliton solution for $N\geq 3$ by taking
\begin{eqnarray}
F=\sum_{1\leq k_1<k_2<\cdots<k_{N-1}\leq N} e^{\Lambda_{k_1,F}+\Lambda_{k_2,F}+\cdots+\Lambda_{k_{N-1},F}},\\
 G=e^{\sum_{i=1}^N\Lambda_{i,G}}.
\end{eqnarray}
For $N=4$, we get the following expressions for the superfields $F$ and $G$ :
\begin{eqnarray}
F=e^{\Lambda_{1,F}+\Lambda_{2,F}+\Lambda_{3,F}}+e^{\Lambda_{1,F}+\Lambda_{2,F}+\Lambda_{4,F}}+e^{\Lambda_{1,F}+\Lambda_{3,F}+\Lambda_{4,F}}+e^{\Lambda_{2,F}+\Lambda_{3,F}+\Lambda_{4,F}},\nonumber\\
G=e^{\Lambda_{1,G}+\Lambda_{2,G}+\Lambda_{3,G}+\Lambda_{4,G}}.
\end{eqnarray}

\section{The supersymmetric Burgers equation}

To show the universality of this new approach, we apply it to the $N=2$ supersymmetric extension of the potential Burgers equation \cite{Hussin1}
\begin{equation}
C_t=D_1D_2C_x+\frac12 C_x^2,
\label{Burgers}
\end{equation}
where $C=u+\theta_1\xi_1+\theta_2\xi_2+\theta_1\theta_2v$ is an even superfield of $\mathbb{R}^{2\vert 2}$. As shown in section 2, we can write $D_1D_2C_x$ as $W_{xx}$. This makes possible the use of the supersymmetric Bell polynomials. With this transformation, equation (\ref{Burgers}) takes the form
\begin{equation}
C_t=W_{xx}+\frac12 C_x^2.
\label{GoogBurgers}
\end{equation}
The above equation can be re-written using the binary Bell polynomials as
\begin{equation}
2\alpha \mathcal{Y}_t(\alpha C,2\alpha^2W)-\mathcal{Y}_{xx}(\alpha C,2\alpha^2 W)=0,
\label{BellBurgers}
\end{equation}
where $\alpha\neq 0$ is an arbitrary even constant. To link (\ref{BellBurgers}) to the Hirota bilinear formalism, we cast the following change of variables:
\begin{equation}
\alpha C=\ln\left(\frac{F}{G}\right) \quad \mbox{and} \quad 2\alpha^2 W=\ln (FG),
\end{equation}
where $F$ and $G$ are two even superfields of $\mathbb{R}^{2\vert 2}$. In this case, equation (\ref{BellBurgers}) becomes
\begin{equation}
(2\alpha \mathcal{D}_t-\mathcal{D}_x^2)(F . G)=0.
\end{equation}
Furthermore, the relation $W_{xx}=D_1D_2C_x$ can be re-written after one integration with respect to $x$ as 
\begin{equation}
\mathcal{S}_1(\mathcal{S}_1-2\alpha S_2)(F.F)=0 \quad \mbox{and} \quad \mathcal{S}_1(\mathcal{S}_1+2\alpha S_2)(G.G)=0.
\end{equation}

We thus get the $N=2$ supersymmetric Hirota bilinear representation of the potential Burgers equation given as
\begin{equation}
\left\{\begin{array}{ll}
(2\alpha\mathcal{D}_t-\mathcal{D}_x^2)(F.G)&=0,\\
\mathcal{S}_1(\mathcal{S}_1-2\alpha\mathcal{S}_2)(F.F)&=0,\\
\mathcal{S}_1(\mathcal{S}_1+2\alpha\mathcal{S}_2)(G.G)&=0.
\end{array}
\right.
\label{HirotaBurgers}
\end{equation}

To solve these equations, we use the following three ansatz:
\begin{enumerate}
\item \textbf{Ansatz 1:} $F=e^{\Lambda_F}$ and $G=1$,
\item \textbf{Ansatz 2:} $F=1$ and $G=e^{\Lambda_G}$,
\item \textbf{Ansatz 3:} $F=e^{\Lambda_F}$ and $G=e^{\Lambda_G}$,
\end{enumerate}
where 
\begin{equation}
\Lambda_{\mu}=\kappa_{\mu}x+\omega_{\mu} t+\theta_1\eta_{1,\mu}+\theta_2\eta_{2,\mu}+\theta_1\theta_2m_{12,\mu},\quad \mu=F,G.
\end{equation}
 For ansatz 1, it is sufficient to impose $(2\alpha\partial_t-\partial_x^2)\Lambda_{F}=(D_1-2\alpha D_2)\Lambda_F=0$ and, for ansatz 2, $(2\alpha\partial_t-\partial_x^2)\Lambda_{G}=(D_1+2\alpha D_2)\Lambda_G=0$.
 
 It is easily shown that the first two ansatz are satisfied for 
 \begin{equation}
 \omega_{\mu}=-i\kappa_{\mu}^2, \quad \eta_{2,F}=-i\eta_{1,F},\quad \eta_{2,G}=i\eta_{1,G}\quad \mbox{and}\quad 2\alpha=i.
 \end{equation}
The third ansatz models, in a certain sense, interactions between exponentials of $F$ and $G$. Ansatz 3 imposes that $\kappa_G=\kappa_F=\kappa$. Hence, we take
\begin{eqnarray}
\Lambda_F=\lambda+(\theta_1-i\theta_2)\eta_F+i\theta_1\theta_2\kappa,\\
\Lambda_G=\lambda+(\theta_1+i\theta_2)\eta_G-i\theta_1\theta_2\kappa,
\end{eqnarray}
where $\lambda=\kappa x-i\kappa^2t$ and $\eta_{\mu}$ are odd constants for $\mu=F,G$.

We deduce that 
\begin{equation}
F=a_{0,F}+a_{1,F}e^{\Lambda_F}\quad \mbox{and}\quad G=a_{0,G}+a_{1,G}e^{\Lambda_G}
\end{equation}
 are solutions of system (\ref{HirotaBurgers}) for all choices of $a_{\mu,\nu}$.
 
 For example, let us take $F=1$ and $G=1+e^{\lambda_G}$. We thus get
 \begin{equation}
 C=2i\ln(1+e^{\lambda})+(\theta_1+i\theta_2)\left(\frac{2i\eta_Ge^{\lambda}}{1+e^{\lambda}}\right)+\theta_1\theta_2\frac{2\kappa e^{\lambda}}{1+e^{\lambda}}.
 \end{equation}
\section{Conclusion}
In this paper, we have propose for the first time a $N=2$ supersymmetric Hirota bilinear representation of the KdV equation for $a=1$ and of the potential Burgers equation. We have expose solitary wave solutions and a $N$-soliton solution.

An ongoing project is to extract similarity solutions using the bilinear Hirota representation and give $N=2$ Hirota bilinear equations for the supersymmetric KdV equation in the other completely integrable cases $a=-2$ and $a=4$.

\section{Acknowledgement}

The author would like to thank the referees for its time and comments.

\end{document}